\documentclass[sigconf, language=french, language=english]{acmart} 

\usepackage[normalem]{ulem} 

\usepackage{xcolor}

\def\eg{{e.g.,} }
\def\cad{càd}
\def\ex{{ex~:} }
\def\pex{{p.~ex.,} }

\setcitestyle{notesep={: }}

\usepackage{eurosym}

\DeclareTextSymbol{\deg}{T1}{6}
\DeclareTextSymbol{\deg}{OT1}{23}

\setcitestyle{nocompress}

\usepackage{hyperref}

\addto\extrasenglish{

}

\addto\extrasfrench{

}

\usepackage{subcaption}

\newcommand{\killpunct}[1]{}

\usepackage{colortbl} 

\usepackage{enumitem}

\usepackage{printlen}

\def\eg{{e.g.,} }

\DeclareTextSymbol{\deg}{T1}{6}
\DeclareTextSymbol{\deg}{OT1}{23}

\newcommand{\g}[1]{``#1''}

\usepackage{amsmath}

\def\myshorttitle{Observing Interaction Rather Than Interfaces}
\def\mytitle{Observing Interaction Rather Than Interfaces}

\def\mytitleFR{Observer l'interaction plutôt que les interfaces}

\def\mykeywords{research methodology, user-centered design, applicative cases, experimental HCI, empirical studies, user studies, user experiments, serendipity, opportunistic research, replication, results replication, replication studies, property studies, interaction loops, interaction prisms, interactional properties} 

\def\mykeywordsFR{méthodologie de recherche, conception centrée utilisateur, cas applicatifs, IHM expérimentale, études empiriques, études utilisateurs, expérimentations utilisateurs, sérendipité, recherche opportuniste, réplication, réplication de résultats, études réplicatives, études de propriétés, boucles interactionnelles, prismes interactionnels, propriétés interactionnelles}


\AtBeginDocument{}

\newcommand{\venue}{IHM '25}
\setcopyright{none}
\acmConference[\venue{}]{36\textsuperscript{e} conférence Francophone sur l'Interaction Humain-Machine}{Novembre 4--7, 2025}{Toulouse, France}
\acmBooktitle{\venue{}~: Actes étendus de la 36\textsuperscript{e} conférence Francophone sur l'Interaction Humain-Machine, Novembre 4--7, 2025, Toulouse, France}
\acmPrice{15.00}
\acmISBN{}
\acmDOI{}

\begin{document}

\title[\myshorttitle]{\mytitle}
\translatedtitle{french}{\mytitleFR}

\author{Guillaume Rivière}
\email{g.riviere@estia.fr}
\orcid{0000-0001-8390-9751}
\affiliation{\institution{Univ. Bordeaux, ESTIA-Institute of Technology, EstiaR}
  \postcode{F-64210}
  \city{Bidart}
  \country{France}
}
\renewcommand{\shortauthors}{Riviere}

\begin{abstract}
The science of Human--Computer Interaction (HCI) is populated by isolated empirical findings, often tied to specific technologies, designs, and tasks.
This situation probably lies in observing the wrong object of study, that is to say, observing interfaces rather than interaction.
This paper proposes an experimental methodology, powered by a research methodology, that enables tackling the ambition of observing interaction (rather than interfaces).
These observations are done during the treatment of applicative cases, allowing to generate and replicate results covering various experimental conditions, expressed from the need of end users and the evolution of technologies.
Performing these observations when developing applicative prototypes illustrating novel technologies' utility allows, in the same time, to benefit from an optimization of these prototypes to better accomplish end users tasks.
This paper depicts a long term research direction, from generating the initial observations of interaction properties and their replication, to their integration, that would then lead to exploring the possible relations existing between those properties, to end toward the description of human--computer interaction's physics.
\end{abstract}
 
\begin{translatedabstract}{french}
La science de l'Interaction Humain-Machine (IHM) revêt beaucoup de résultats empiriques isolés, souvent liés à des technologies, des conceptions et des tâches spécifiques. Cette situation tient certainement de l'observation du mauvais objet d'étude, soit -- en d'autres termes -- d'observer les interfaces plutôt que l'interaction. Le présent article propose une méthodologie expérimentale, s'inscrivant dans une méthodologie de recherche, permettant de répondre à l'ambition d'observer l'interaction (plutôt que les interfaces). Les observations sont conduites lors du traitement de cas applicatifs, permettant de générer et répliquer des résultats en couvrant des conditions expérimentales variées, issues des besoins des personnes utilisatrices finales et de l'évolution des technologies. Conduire ces observations lors du développement de prototypes applicatifs illustrant l'utilité des nouvelles technologies permet, dans le même temps, d'en tirer le bénéfice par une optimisation de ces prototypes pour un meilleur accomplissement des tâches des personnes utilisatrices finales. Cet article dépeint une direction de recherche à long terme, allant de la génération d'observations initiales de propriétés interactionnelles et à leur réplication, à leur intégration, puis qui conduirait à l'exploration des possibles relations qui existeraient entre ces propriétés pour aller vers la description d'une physique des interactions humain-machine.
\end{translatedabstract}

\translatedkeywords{french}{\mykeywordsFR}

\maketitle

\selectlanguage{french}

\section{Introduction}

Déterminer des approches scientifiques au domaine de l'Interaction Humain-Machine (IHM) est toujours une préoccupation centrale depuis quatre décennies \cite{reeves2015science}~: depuis les échanges académiques entre Newell, Card, Carroll et Campbell \cite{carroll1986softening,newell1985prospects,newell1987straightening} du milieu des années 1980 et jusqu'à de plus récentes discussions (\eg{} \cite{carroll1997hci,carroll2010conceptualizing,dix2010hci,greenberg1992weak,john1989cumulating,john1990conceptions,kostakos2015hole,olsen2007evaluating,reeves2015science,rogers2004new,sutcliffe1991trial}).
Les principaux questionnements portent sur le manque de théories \cite{greenberg1992weak}, d'hypothèses risquées \cite{greenberg1992weak,greenberg2008usability} et de réplication des résultats empiriques \cite{greenberg1992weak}, conduisant à surgénéraliser des résultats isolés \cite{greenberg1992weak}. Encore des travaux récents interrogent sur la génération de modèles théoriques \cite{mbl2021generative}. En outre, la course à la publication de démonstrateurs originaux résulte en d'importants efforts en ingénierie pour créer de nouvelles interfaces surpassant les précédentes, qui restent malencontreusement trop éloignés et distincts des efforts scientifiques à vouloir comprendre, expliquer et répliquer les observations empiriques \cite{greenberg1992weak,greenberg2008usability} (des ateliers internationaux ont même été spécifiquement consacrés à cette problématique de la réplication \cite{wilson2011replichi,wilson2012replichisig,wilson2013replichi,wilson2014replichi2}).
Pourtant, sans théories, l'ingénierie en IHM est condamnée à tâtonner avec des résultats empiriques ne survivant pas aux conditions expérimentales dans lesquelles ils ont été observés \cite{greenberg1992weak}.

Aujourd'hui, par exemple, nous bénéficions tous -- dans la recherche, dans l'industrie et dans le quotidien -- des critères ergonomiques \cite{bastien1993criteria}~: ces règles guident, entre autres, la conception de menus, de sites web et d'applications ayant des longueurs et des profondeurs de navigation raisonnables. Cependant, beaucoup de prototypes de recherche continuent de reposer sur l'intuition et des principes d'ingénierie \cite{greenberg1992weak} basés sur des observations isolées. De plus, ces observations isolées apparaissent dans des conditions particulières \cite{greenberg1992weak}, ne survivant pas aux prototypes, aux technologies et aux tâches impliquées lors de l'étude où elles sont apparues.
L'ère de l'informatique ubiquitaire \cite{weiser1991century,weiser1999century} a très certainement exacerbé cette situation, pour avoir incité à trouver et explorer les multiples genres d'interfaces humain-machine offerts et leurs possibles applications.
Contrairement aux précédentes interfaces par lignes de commandes (CLIs) et aux interfaces graphiques (GUIs), l'ère de l'informatique ubiquitaire \cite{weiser1991century,weiser1999century} a conduit à une multitude de formes physiques d'interfaces et de techniques d'interaction (\cad{} les interfaces Post-WIMP \cite{mbl2000instrumental}, recourant à l'interaction multimodale \cite{nigay1993multimodal}, \pex{} impliquant plusieurs sens humains en entrée et en sortie des systèmes, des nombres divers de personnes utilisatrices et des contextes d'utilisation variés). L'exploration de l'espace de conception qui en a résulté a engendré profusion de spécimens. Établir les théories pertinentes expliquant et prédisant les phénomènes interactionnels impliqués dans tous ces différents genres et spécimens d'interfaces ubiquitaires est ainsi un travail colossal, demandant alors de nombreux efforts sur les théories en IHM de la part des communautés de recherche. Pour autant, formuler des hypothèses et transformer les résultats cumulés en règles générales est valable et nécessaire car, sinon, l'IHM continuera de produire sans cesse des observations sans comprendre pourquoi les interfaces humain-machine sont bien ou mal conçues, ni comment les améliorer. Le présent article repose sur la conjecture que la persistance de cette situation est indépendante d'un manque de volonté, et résulte plutôt d'un manque de méthodologie de recherche adaptée, ainsi que d'observations portant sur le mauvais objet d'étude. Une façon de s'en convaincre sera, tout comme pour le déchiffrement de l'écriture Maya\footnote{Le déchiffrement de l'écriture Maya fut supposé correct enfin une fois que les décodages de textes se sont alors accrus rapidement.}, l'accumulation rapide de nouvelles observations variées, puis l'apparition de nouveaux résultats, règles ou lois.

Le présent article suggère une méthodologie de recherche duale, d'optimisation et d'apprentissage, qui s'appuie sur les efforts en ingénierie des communautés de recherche en IHM pour alimenter en même temps les efforts en théorie des IHMs. Durant la première phase, des efforts en ingénierie approfondissent leur examen des prototypes et des tâches provenant de cas applicatifs (\cad{} répondant à des besoins réels et obtenus, par exemple, par conception centrée utilisation) en y adjoignant des \g{études de propriétés} \cite{riviere2025initiating}. Ces études de propriétés interactionnelles permettent de produire de nouvelles observations par variation de paramètres du prototype et, éventuellement, de la difficulté de la tâche, mais de manière calibrée~: la variation d'un seul paramètre est possible à la fois, en utilisant un outil imaginaire appelé \g{prisme interactionnel}. L'idée est, comme pour l'étude de rayons invisibles en optique ondulatoire, de révéler les boucles d'interaction par leur diffraction en un point. Cette diffraction permet une double contribution. Premièrement, elle permet de guider l'amélioration du prototype étudié en optimisant la valeur d'un de ses paramètres. Deuxièmement, elle produit une nouvelle observation calibrée d'un paramètre isolé, qui pourra alimenter la confirmation, l'extension ou la création de résultats issus d'observations répliquées. En revanche, parce que les cas applicatifs sont pris comme point de départ, la conception des prototypes et le choix des paramètres à optimiser doivent uniquement se voir guidés par une réponse adaptée aux besoins des futures personnes utilisatrices, c'est-à-dire sans se préoccuper du soucis de devoir alimenter une théorie en cours d'élaboration. De la sorte, la présente méthodologie de recherche repose pleinement sur le principe de sérendipité. La deuxième phase, quant à elle, consiste à généraliser les observations accumulées de manière opportuniste par diverses études de propriétés, en confortant, consolidant ou faisant émerger de nouvelles théories. Pour ce faire, deux modes opératoires sont envisageables~: les observations cumulées sont transformées en un résultat qui sera validé par une expérimentation de vérification d'hypothèse, ou encore, elles sont combinées et pondérées par une méta-analyse.

Ainsi, cette méthodologie de recherche, nommée \g{recherche surjective}, partant de multiples cas applicatifs pertinents pour aller vers des théories, s'inscrit dans une démarche de recherche inductive basée sur la réplication. La réplication sur des cas applicatifs est bénéfique de deux façons. Tout d'abord, cela contribue à préciser les conditions dans lesquelles des résultats connus sont valides et à éliminer ceux ne pouvant être répliqués dans d'autres conditions que celles de l'expérimentation originale \cite{hornbaek2014once}. Ainsi, au-delà de l'optimisation de prototypes, les études de propriétés sur de nouveaux cas applicatifs contribuent à la réplication de résultats connus de sorte à pouvoir vérifier si les hypothèses théoriques continuent de se vérifier dans de nouveaux contextes et de nouvelles conditions expérimentales (\cad{} de la réplication partielle ou conceptuelle \cite{hornbaek2014once}). En revanche, cette méthode reste distincte des démarches de reproduction de protocoles et de conditions expérimentales visant à vérifier la validité de travaux déjà publiés \cite{shepperd2018replication} (\cad{} de la réplication stricte \cite{hornbaek2014once}).

Malheureusement, les études de simple réplication en IHM sont peu récompensées, peu reconnues et peu publiées \cite{hornbaek2014once}. Cependant, la recherche surjective se veut aussi un remède vertueux pour accroître la publication des études de réplication. En effet, mener des études de propriété sur des cas applicatifs demandera un budget et un effort de recherche et de développement qui seront en parti amortis pour avoir déjà analysé les tâches des personnes utilisatrices finales, puis développé ou mobilisé des prototypes dans le but de répondre à un besoin de l'industrie ou du quotidien. Surtout, greffer des études de propriétés aux études d'utilisation et aux études comparatives, usuelles lors du traitement de cas applicatifs, seront à la fois récompensés par l'amélioration d'un paramètre du prototype spécifiquement conçu, par une meilleure justification des choix de conception et par une meilleure compréhension des raisons du bon fonctionnement du prototype. Ainsi, d'un côté, ces études de propriétés permettent de contribuer à l'effort de développement technologique déjà prépondérant au sein des communautés en IHM. Aussi, d'un autre côté, elles permettent à la communauté en IHM de générer des réplications de résultats connus ou de nouveaux résultats (qui seront à répliquer), contribuant ainsi à l'effort scientifique de la communauté en amortissant une partie de l'effort de développement et financement par réponse à des besoins de personnes utilisatrices finales. Cette méthode permet donc d'enrichir de simples cas applicatifs par greffe d'études à vocation réplicatives et d'obtenir ainsi une double reconnaissance technologique et scientifique facilitant ainsi leur publication.

L'article commence par contextualiser la présente recherche, puis énonce sa proposition, avant d'ensuite planifier les futures recherches, puis de conclure.

\section{Contextualisation}

Parmi les études d'utilisation actuellement conduites, nous pouvons distinguer les conditions expérimentales et objectifs expérimentaux en au moins quatre sortes.

\begin{enumerate}
  \item[(S1)] Les études prospectant ou vérifiant une loi (\ex{} \cite{fitts1954law}) ou un principe (\ex{} \cite{guiard1987asymmetric});
  \item[(S2)] Les études testant un prototype applicatif répondant à un besoin spécifique (\ex{} \cite{legardeur2004eskua,reuter2007archeotui});
  \item[(S3)] Les études comparant des techniques d'interaction (\ex{} \cite{bailly2007wave});
  \item[(S4)] Les études comparant des technologies et des prototypes (\ex{} \cite{froehlich2006globe,hachet2003cat}).
\end{enumerate}

Ces études peuvent également se distinguer par les types d'interfaces qu'elles impliquent. Soit une étude $e$. Soit $R$ l'espace des résultats dans lequel elle génère des observations. Soient $T_i$ des types d'interfaces (\ex{} CLI, GUI, TUI, RA, RV\ldots). Cette étude sera dîte ``simple'' si elle engage une seule condition d'interaction (\ex{} S1, S2) :
\[e : T_a  \longmapsto R\]

Cette étude sera dîte ``plurielle'' si elle engage plusieurs conditions d'interaction. Elle sera ``homogène'' lorsque toutes les conditions d'interaction sont du même type d'interface (\ex{} S3, uniquement la technique d'interaction varie) :
\[e : T_a \times T_a \times T_a \longmapsto R\]

Elle sera dîte ``hétérogène'' lorsque les conditions d'interaction sont de plusieurs types d'interfaces (\ex{} S4) :
\[e : T_a \times T_b \times T_c \longmapsto R\]

Malheureusement, les observations issues d'une étude sont rarement répliquées, faisant de ses résultats la conclusion d'une unique observation, à l'exception de quelques cas, comme celui des études de dispositifs de pointage (le plus souvent universels, telle la souris, mais pas uniquement), via des tâches calibrées (\ex{} \cite{iso:9241-9:2000}), reproduisant, précisant ou revisitant une loi connue (\ex{} la loi de Fitts \cite{blanch2011benchmarking}). Ces rares cas de réplication commencent récemment à faire l'objet de méta-analyses (\ex{} \cite{amini2025systematic}).

Ce décalage entre les efforts en ingénierie et les approches scientifiques de l'IHM a pu être rendu visible par le type des contributions et les taux de réplication. D'une part, la conférence CHI 2016 a publié 546 articles, dont 44,0\% concernaient des études empiriques sur l'utilisation de systèmes interactifs (240 articles) et 24,5\% sur de nouveaux artefacts ou systèmes (134 articles) \cite{wobbrock2016contributions}. Enfin, seulement 4,6\% concernaient des contributions théoriques (25 articles) et 1,8\% des méta-analyses et des revues de la littérature (10 articles) \cite{wobbrock2016contributions}. D'autre part, entre 2008 et 2010, de 891 articles publiés par les conférences CHI et les revues ToCHI, HCI et IJHCS, seulement 3\% s'intéressaient à la réplication de résultats antérieurs (28 articles) \cite{hornbaek2014once}. Depuis ces constatations, il semble qu'à ce jour aucune proposition forte ne soit venue contrecarrer la situation.

L'origine de ce manque de consolidation des résultats par répétition des observations (\cad{}, par mesures répétées) est probablement autre qu'un manque de volonté des communautés de recherche et possiblement celle d'un manque de méthode, permettant la coordination d'un effort collectif répartissant la charge de travail à l'échelle des communautés de recherche, tant que celui d'un objet d'étude adéquat, se prêtant à des observations indépendantes des technologies et prototypes utilisés lors des observations, tant que celui du croisement des compétences entre les équipes et institutions développant des technologies, des techniques d'interaction ou des prototypes applicatifs, ainsi que celles formalisant les concepts ou découvrant les principes.

S'infiltrant dans cette veine, la section suivante propose de prendre l'interaction comme objet d'étude, plutôt que les interfaces, procédant d'un glissement dans la méthode expérimentale et la façon d'interpréter les observations, impactant également la façon de construire des projets de recherche viables tant scientifiquement qu'économiquement parlant (\cad{} une approche inductive que nous inscrivons dans un type d'approche que nous qualifions de ``recherche surjective'').

\section{Proposition}

La proposition porte en premier lieu sur l'objet d'étude, en observant l'interaction plutôt que les interfaces, devenant ainsi indépendant des technologies, prototypes et techniques d'interaction sollicitées. Le terme ``interaction'' a de plus en plus reçu l'attention des recherches en IHM au fil du temps \cite{hornbaek2019mean}, par un glissement de la perspective prise sur le champ d'étude, pour passer de la création ``d'interfaces humain-machine'' à la création ``d'interactions''  \cite{mbl2004designing}. Or, la perspective prise pour un objet d'étude est conséquente dans le sens où elle conditionne son observation et son interprétation, mais aussi les moyens d'évaluation et de conception \cite{hornbaek2017interaction}. Une récente définition de ce terme indique que  ``l'interaction`` survient lorsque ``\textit{deux entités déterminent mutuellement leurs comportements respectifs au cours du temps}'' et que, en l'occurrence, ces entités sont un système interactif et une personne utilisatrice \cite{hornbaek2017interaction}. Une des façons de voir cette interaction est celui d'un ``\textit{dialogue composé d'échanges cycliques, passant par des canaux d'entrées et de sorties (du point de vue du système interactif) ou des canaux d'action et de perception (du point de vue de la personne utilisatrice)}'' \cite{hornbaek2017interaction}.

Cette perspective prise sur l'interaction, vue comme un dialogue, permet alors une caractérisation particulière. Ainsi, selon la théorie de l'action, ces échanges cycliques consistent, pour la personne utilisatrice, en sept étapes \cite{norman2002design}. De plus, plusieurs distances apparaissent de part et d'autre des cycles, au sein du gouffre de l'exécution et du gouffre de l'évaluation, et sont à réduire pour obtenir une meilleure interaction \cite{norman2002design}.
Enfin, comme ces échanges surviennent de manière dynamique, au cours du temps, l'interaction peut se voir comme un phénomène, se traduisant par une alternance d'échanges au cours de cycles successifs. Dans la littérature, les différents cycles possibles avec une interface numérique sont le plus souvent retrouvés sous l'appellation de ``boucles'' voire de ``boucles de rétroaction''. Étudier les boucles d'interaction est donc un moyen d'étudier l'interaction.

\subsection{Révéler l'interaction}

L'idée est, comme pour l'étude de rayons invisibles en optique ondulatoire, de rendre apparentes les boucles d'interaction du dialogue humain-machine par leur altération en un point. Les incidences observées résultent donc de l'altération en ce point. Cette altération, que nous appelons ``diffraction'', consiste à faire varier un paramètre de l'interface en lui donnant successivement plusieurs valeurs. La diffraction devient visible à condition que le paramètre et les valeurs choisies soient suffisamment impactantes. Prolongeant la métaphore de l'optique ondulatoire, nous symbolisons l'action de diffraction d'une boucle d'interaction en un point par la pose d'un outil imaginaire de ``prisme interactionnel''. D'un point de vue de la statistique, un prisme interactionnel décrit simplement une variable indépendante. En revanche, le terme de prisme interactionnel se veut spécifique aux boucles d'interaction. De plus, l'utilisation d'une terminologie particulière permet d'orienter l'interprétation des résultats. Ainsi, la diffraction de boucles d'interaction permet de s'imposer une certaine prise de vue~: observer l'interaction plutôt que les interfaces. Pour ce faire, l'observation de l'interaction par diffraction est simplement une façon de procéder à une étude d'utilisation et une façon d'en interpréter les résultats (en considérant avoir altéré une boucle).

Les protocoles expérimentaux de ces études prendront alors comme variable indépendante un (et un seul) paramètre $p$ de l'interface\footnote{Les variables indépendantes sont communes à tout protocole expérimental, mais une variable indépendante n'est pas forcément un paramètre de l'interface.}, qui prendra plusieurs valeurs $v_i$, et feront éventuellement varier le niveau de tâche (\ex{} $t_1$ de bas niveau et $t_2$ de haut niveau) pour l'accomplissement de tâches représentatives du besoin des personnes utilisatrices finales, exécutées sur des prototypes à vocation applicative. Ces études -- que nous nommons ``études de propriétés'' -- sont homogènes :
\[e(p, t_1) : T_a(v_1) \times T_a(v_2) \times T_a(v_3) \longmapsto R\]
\[e(p, t_2) : T_a(v_1) \times T_a(v_2) \times T_a(v_3) \longmapsto R\]

\subsection{Interpréter les observations}

Nous définissons l'effet d'un paramètre sur l'interaction, observé lors d'une diffraction, comme une ``propriété interactionnelle''. Typiquement, les propriétés interactionnelles expliquent comment les caractéristiques des interfaces humain-machine affectent l'interaction à cause des capacités, facultés et aptitudes humaines. Ces propriétés sont des assertions pouvant être vraies ou fausses (au même sens que les propriétés en mathématiques), décrivant des relations d'ordre entre les valeurs du paramètre, pour des caractéristiques de l'interaction (\ex{} performance, préférences, \ldots), et précisant les conditions dans lesquelles ces relations s'exercent ou s'estompent (\pex{} selon le niveau ou le type de tâche). Nous appelons dès lors ``études de propriétés'' les études d'utilisation précédemment décrites procédant par diffraction selon un paramètre.

\subsection{Explorer les boucles d'interaction}

Des modèles d'interaction ont déjà permis de mieux comprendre les boucles d'interaction, en distinguant leur phase d'implication dans l'interaction (\ex{} \cite{ishii2008beyond,mbl2000instrumental}), en les décomposant en canaux directionnels (entrées et sorties, ou actions et de perceptions, selon le point de vue pris \cite{hornbaek2017interaction}) et en les segmentant en plusieurs étapes (\ex{} \cite{dumas2009multimodal}). Cette compréhension théorique préalable pourra autant aider dans l'interprétation des observations, qu'éventuellement dans l'exploration systématique de l'interaction par diffraction des boucles selon leurs phases d'implication et par placement de prismes en leurs différents canaux directionnels et différentes étapes.

Une fois que les boucles auront été suffisamment explorées et de propriétés interactionnelles pu être établies (par une meilleure connaissance de l'effet individuel de l'altération des boucles en chacun de leurs points), des diffractions multiples pourront alors s'envisager pour établir les possibles relations qui peuvent exister entre les propriétés interactionnelles (\pex{} fonction des paramètres impliqués, des caractéristiques ciblées, ou des boucles concernées et de leurs points). Cette exploration avancée pourra s'accomplir par diffraction en de multiples points à la fois, ou de multiples boucles en même temps, la connaissance préalable des propriétés impliquées permettant et facilitant l'interprétation des observations nouvellement recueillies. La connaissance de ces relations permettrait alors d'envisager l'ébauche d'une physique des interactions humain-machine.

\subsection{Provoquer le phénomène}

Pour pouvoir observer l'interaction, un phénomène doit se produire dynamiquement (\cad{} des échanges entre personnes utilisatrices et systèmes interactifs). Nous pouvons trouver ce phénomène en situation écologique, alors que des personnes utilisatrices interagissent dans leur quotidien. Instaurer un protocole de diffraction dans ces conditions d'interaction reste néanmoins long et fastidieux relativement au nombre d'observations à entreprendre pour une exploration et une découverte des boucles d'interaction en profondeur.

L'interaction peut aussi se provoquer en conditions artificielles. Par exemple, les boucles d'interactions pourraient s'explorer de manière systématique en isolant et focalisant sur chacun de leurs points. Il serait alors demandé aux personnes utilisatrices d'exécuter des tâches très spécifiques, se limitant par exemple à une seule action. Suivre une telle stratégie a déjà permis de mieux comprendre les actions de pointage et d'établir la loi de Fitts \cite{fitts1954law}. Cependant, l'inconvénient de cette stratégie est que le choix des points à diffracter reposerait uniquement sur des bases théoriques et le bornerait aux modèles d'interaction à disposition. De plus, l'interaction est un processus complexe, incluant des aspects tant manipulatoires et opératoires que cognitifs, tant d'exécution que de raisonnement et de planification \cite{norman2002design} et étudier les boucles autrement que dans leur inclusion dans cette complexité globale, c'est-à-dire en voulant isoler les actions des tâches et sous-tâches qui les sollicitent, manquerait certainement d'en saisir toutes les nuances. Par exemple, un paramètre peut être sensible aux niveaux de tâche et à la caractéristique étudiée (\ex{} \cite{becher2021projectionEN,becher2021projectionFR,couture2008geotui}).

Une deuxième voie pour provoquer le phénomène interactionnel serait de piocher dans un catalogue répertoriant des représentantes des tâches trouvées habituellement dans les applications existantes (\ex{} les boîtes de dialogue \cite{mbl2000instrumental}). Cette stratégie est certainement viable dans le cadre de l'informatique personnelle et mobile, avec des types d'interfaces fixes et des paradigmes d'interfaces connus (\ex{} GUI avec WIMP) et pour des formes d'interfaces universelles (\ex{} écran/clavier/pointage, écran/tactile multipoint). Cependant, la variété des interfaces de l'informatique ubiquitaire, et leur constante évolution, rend potentiellement encore laborieux -- et peu probable -- le succès d'un inventaire systématique des tâches existantes en quelques représentantes. De plus, encore ici, ces tâches devraient pouvoir continuer de rendre compte des raisonnements induits par les tâches des personnes utilisatrices au cours de l'exécution (ce qui peut, somme toute, se simuler par des tâches de distraction, \ex{} \cite{bailly2021exploration,daniel2019cairnform}).

Une troisième voie réside dans la prospection opportuniste de tâches ou sous-tâches issues des besoins des personnes utilisatrices finales. Cette stratégie est plus particulièrement adaptée au cadre de l'informatique ubiquitaire et de la multitude de formes que prennent ses systèmes interactifs et de la variété des tâches concernées, d'autant qu'aucuns paradigmes unificateurs ne guident encore réellement la conception de ces systèmes. L'avantage de cette stratégie est de pouvoir envisager d'observer l'interaction dans sa complexité et dans sa diversité, à condition de pouvoir multiplier les cas applicatifs. En corollaire, l'inconvénient de cette stratégie est donc de nécessiter de rencontrer beaucoup de personnes utilisatrices finales, de recueillir leurs besoins et de développer tout autant de prototypes. La conduite d'un tel chantier semble devoir se dimensionner à l'échelle de communautés de recherche et nécessiter qu'elles puissent se coordonner par des méthodes expérimentales calibrées (études et forme des résultats). Les multiples observations ainsi générées nécessiteront ensuite leur intégration (\ex{} méta-analyses, construction de théories, généralisation). Cette troisième voie est celle que suggère de suivre la présente recherche.

\subsection{Soutenir l'effort de recherche}

Vouloir observer l'interaction au travers d'une diversité de tâches, issues des besoins d'une variété de personnes utilisatrices finales, induit de soutenir un effort de recherche suffisamment ample et prolongé, pour couvrir des cas applicatifs nombreux, ainsi que convoquer les technologies actuelles et à venir, puis récolter et intégrer la pléthore d'observations obtenues. Pour ce faire, cet article appuie l'idée d'un effort de recherche duale, à vocation scientifique et technologique. Démontrer l'utilité des nouvelles technologies pourrait en effet s'accompagner d'études de propriétés, permettant d'optimiser un prototype applicatif en trouvant la meilleure valeur pour un ou plusieurs de ses paramètres. Dans le même temps, générer ces observations va permettre d'émettre ou consolider un résultat théorique futur ou antérieur, participant ainsi à la consolidation des connaissances sur l'interaction. Le recueil, au fil des recherches, de résultats au travers de prototypes variés et de tâches variées permet alors d'observer l'interaction dans des conditions diversifiées. En revanche, le paramètre se devant de rester premièrement déterminé par l'optimisation de la tâche des personnes utilisatrices finales, ce choix ne peut se voir commandé par le besoin d'alimenter un résultat déjà connu ou une théorie en cours de construction. La nécessité demeure, donc, de voir la méthode expérimentale répétée dans de nombreux projets de recherche, de sorte à générer et répliquer les observations suffisantes pour que puissent devenir envisageables l'intégration et la généralisation de résultats (\pex{} se décidant après saturation, convergence ou stabilisation des observations).

Nous devons souligner le caractère inductif et opportuniste de la méthodologie de recherche qui est ainsi empruntée. Le choix des paramètres observés sera premièrement guidé par le souhait d'optimisation des prototypes applicatifs pour accomplir les tâches des personnes utilisatrices finales (et non pas par le souhait de construire ou étayer une théorie). Les observations ainsi générées seront donc issues d'applications pertinentes, et les ensembles d'observations qui en résulteront seront tout autant pertinents. Les théories construites à partir de ces ensembles d'observations seront également pertinentes, car déjà impliquées dans de nombreuses applications. Par sa vision ensembliste, nous appelons ``recherche surjective'' cette approche inductive, élaborant des théories construites à partir d'ensembles d'observations pertinentes, remplis de manière opportuniste et systématique à partir d'études calibrées sur des prototypes applicatifs. Dans une phase d'intégration, regrouper des observations dans un même ensemble permettra d'élaborer et établir un résultat (\ex{} par méta-analyses), en précisant les conditions nécessaires à son apparition, puis à le généraliser en l'expliquant par une théorie sous-jacente (typiquement liée aux capacités humaines). En complément, et lorsque pertinent (\pex{} pour établir une loi), des expérimentations pourront éventuellement venir vérifier le résultat en conditions de laboratoire sur des tâches ``neutres'' (\cad{} isolées de tout contexte applicatif). 

Un des atouts de cette méthodologie de recherche duale est de mobiliser des ressources pour servir simultanément le développement technologique et l'accumulation de connaissances. Cette méthodologie est donc fondée scientifiquement et économiquement. D'une part, les démonstrateurs de technologies servent à illustrer les possibles, et leur optimisation les rends plus compétitifs au moment des études comparatives, tout en informant les futurs concepteurs des meilleurs choix à opérer pour la tâche. D'autre part, les observations générées permettent de produire les réplications nécessaires à l'ébauche, la consolidation et la validation de nouvelles propriétés interactionnelles, contribuant ainsi à approfondir la compréhension de l'interaction, et à créer une science des interactions humain-machine basée sur des résultats répliqués.

\section{Planification}

Cette proposition d'observation de l'interaction deviendra fructueuse dès que de premières propriétés interactionnelles auront pu être intégrées et généralisées. Cette section pose les jalons qui bornent la route qui nous y conduira.

L'établissement d'un résultat repose sur des observations initiales et un nombre suffisant de réplications. La méthode expérimentale a déjà été mise à l'\oe{}uvre lors de l'étude de cinq prototypes applicatifs (pour les géosciences \cite{couture2008geotui,riviere2009phd}, l'archéologie \cite{reuter2010archeotui,riviere2010activation}, les énergies renouvelables \cite{daniel2019cairnform}, la chirurgie \cite{bailly2021exploration,bailly2020phdthesis} et les travaux publics \cite{becher2021projectionEN,becher2021projectionFR}) et a généré des résultats orthogonaux entre eux, ne se regroupant ni par le paramètre étudié, ni sous une théorie commune. La méthode expérimentale devra donc encore être mise en \oe{}uvre un grand nombre de fois, et sur des prototypes applicatifs variés, avant de voir apparaître les premières opportunités d'intégration de résultats. Le nombre de répétitions nécessaires à l'intégration d'un résultat est difficile à anticiper. Certains résultats, dont la formulation et les conditions d'apparition seront simples, nécessiteront probablement uniquement quelques observations. D'autres résultats, de formulation plus complexe ou dont les conditions d'apparition dépendent de nombreux facteurs, nécessiteront certainement bien plus d'observations.

Une fois que les boucles d'interaction seront décrites de manière encore plus détaillée, et que des propriétés interactionnelles permettront de connaître les incidences en chacun de leurs points, il deviendra possible d'explorer empiriquement et établir les relations qui peuvent résider entre ces propriétés (par diffractions multiples), de sorte à décrire une possible physique interactionnelle.

En outre, la définition actuelle de propriété interactionnelle, ainsi que la méthode de révélation associée, tend à se concentrer sur les propriétés ergonomiques de l'interaction. Un prolongement de la présente recherche serait donc de considérer aussi l'interaction, par exemple, visant à faire vivre une expérience et de déterminer si l'altération de boucles, la réplication de résultats et la conduite de méta-analyses restent alors encore pertinentes.

Ces recherches bénéficieront de formalismes permettant d'exprimer les résultats observés \cite{riviere2025initiating} et de décrire leurs conditions d'émergence \cite{riviere2025initiating}, qui pourront autant apparaître dans les publications que se voir répertoriés dans des catalogues (\pex{} par une description précise des protocoles expérimentaux comme permis par les outils interactifs Touchstone \cite{mackay2007touchstone} et Touchstone2 \cite{eiselmayer2019touchstone2}, sous forme d'une syntaxe XML \cite{mackay2007touchstone} ou d'un langage \cite{eiselmayer2019touchstone2}).

Ces recherches bénéficieront aussi d'avoir mieux cartographié, compris et normalisé les boucles d'interaction impliquées dans les différents genres d'interfaces humain-machine, de sorte à pouvoir raffiner la description des conditions expérimentales et des propriétés interactionnelles. Cette compréhension pourra autant s'affiner par les théories sur les capacités humaines (\ex{} \cite{norman2002design}), que par analyse des applicatifs (\ex{} \cite{mbl2000instrumental}), que par synthèse des diffractions et des positions de prismes prises au fil des études de propriétés conduites lors des recherches applicatives en cours. Ces descriptions affinées des boucles aideront autant la préparation de protocoles, que la description et la généralisation des propriétés.

Ces recherches bénéficieront également d'avoir préalablement recensé les méthodes d'intégration de résultats déjà à l'\oe{}uvre dans d'autres sciences (\pex{} méta-analyses, induction et élaboration de théories, généralisation de résultats) et anticiper leur adaptation aux besoins des études de propriétés. La prise en compte de métriques comme la puissance statistique des protocoles impliqués (comme fourni également par l'outil Touchstone2 \cite{eiselmayer2019touchstone2}) pourra contribuer à peser et pondérer entre les différentes observations recueillies. Ces recherches bénéficieront aussi d'avoir pu anticiper les formes possibles des résultats à venir (\pex{} lois, règles, critères) et sur les possibles façons de les synthétiser, ainsi que la forme des possibles relations qui peuvent exister entre les propriétés interactionnelles.

Enfin, si cette méthodologie de recherche réussit effectivement à engendrer l'effort de recherche escompté à l'échelle communautaire, il deviendra éventuellement judicieux d'essaimer la recherche surjective (avec son principe d'études duales et calibrées) pour soutenir la réplication de résultats dans d'autres domaines de recherche.

\section{Conclusion}

Observer l'interaction, plutôt que les interfaces, tient certainement d'un glissement dans la méthodologie de recherche (pour pouvoir alimenter la génération de réplications) et dans la méthodologie expérimentale (pour révéler les incidences de l'altération de paramètres et en interpréter les observations). Sur ces bases peuvent s'envisager la synthèse de résultats à partir d'observations répliquées, ainsi que s'entrevoir la construction d'une physique interactionnelle décrivant le comportement de l'interaction.

\selectlanguage{english}

\bibliographystyle{ACM-Reference-Format}
\bibliography{paper.bib} 

\end{document}